\documentclass{jetpl}
\twocolumn

\lat

\usepackage{amsmath}
\usepackage{graphicx}

\title{Di-pion emission in heavy quarkonia decays.}

\rtitle{Di-pion emission in heavy quarkonia decays}

\sodtitle{Di-pion emission in heavy quarkonia decays}

\author{Yu.A.Simonov}
\address{ State Research
Center\\Institute of Theoretical and Experimental Physics, \\
Moscow, 117218 Russia}
\newcommand{\beq}{\begin{eqnarray}}
 \newcommand{\eeq}{\end{eqnarray}}
\newcommand{\be}{\begin{equation}}
 \newcommand{\ee}{\end{equation}}

\newcommand{\begat}{\begin{gathered}}
 \newcommand{\eegat}{\end{gathered}}
 \def\la{\mathrel{\mathpalette\fun <}}

\def\fun#1#2{\lower3.6pt\vbox{\baselineskip0pt\lineskip.9pt
\ialign{$\mathsurround=0pt#1\hfil ##\hfil$\crcr#2\crcr\sim\crcr}}}

\newcommand{{\SD}}{\rm SD}

\newcommand{\ver}{\mbox{\boldmath${\rm r}$}}

\newcommand{\vep}{\mbox{\boldmath${\rm p}$}}
\newcommand{\veK}{\mbox{\boldmath${\rm K}$}}

\newcommand{\vek}{\mbox{\boldmath${\rm k}$}}

\newcommand{\lan}{\langle}
\newcommand{\ran}{\rangle}

\abstract{

The di-pion spectrum for  the $\Upsilon(nS) \to \Upsilon (n'S))$
transition with $n\leq 4$  has the form $\frac{dw}{dq}\sim $
(phase space) $ |\eta-x|^2$, with $x=\frac{q^2-4m^2_\pi}{(\Delta
M)^2 -4 m^2_\pi}< q^2 \equiv M^2_{\pi\pi}, $ and $\Delta M=M(nS)
-M(n'S)$. The parameter $\eta $ is calculated and the spectrum is
shown to reproduce the experimental data for all 3 types of
decays: $3\to 1, 2\to 1$ and $3\to 2$ with $\eta \approx 0.5; 0$,
and $-3$, respectively.}


\begin{document}

\maketitle

1. The  di-pion decays of heavy quarkonia are studied for the last
three decades, (see \cite{1}, \cite{2} for references and
discussion). The first observed process $V(2S) \to V(1S) \pi\pi,
V=\psi$ or $\Upsilon$ yields a simple di-pion spectrum, with the
amplitude $M\sim (q^2-4m^2_\pi),$  and theoretical methods have
been envisaged (see \cite{2} for a review  and references), based
on PCAC and the multipole gluon field expansion (MGFE). However,
the $3\to 1 $ and $3\to 2$ di-pion transitions in bottomonium,
observed by CLEO \cite{3}, show two other types of spectra: a
double peaked spectrum for ($3\to 1$) and shallow form for ($3\to
2)$. Numerous  modifications of MGFE and additional models were
suggested  \cite{1,2}, without, however, a unique physical picture
for all 3 types of decays. It is a purpose of this short note to
sketch the general mechanism of di-pion transitions, leaving
details to the full text in \cite{1}.

2.  Our  starting point is the Field Correlator  Method (FCM)
\cite{4}, which is based on the use of Gaussian (quadratic) field
correlators, applicable, in contrast  to MGFE, also for systems of
large size, $R\gg \lambda$ , where $\lambda \la 0.2$ fm is the
vacuum correlation   length \cite{4}. As a result all large
systems display linear confinement and the string tension $\sigma$
defines the  dynamics instead of gluon condensates as in MGFE. It
is important    that all decaying $X(n)$ states have $R\geq 0.4$
fm $>\lambda$, and therefore should be treated in FCM rather than
in MGFE approach.

The pion emission in FCM comes  from  the light quark loop and  is
decribed by the quark-pion Lagrangian \cite{5} \be S_{QM} =- i
\int d^4 x \bar \psi (x) M_{br} \hat U(x) \psi (x)\label{1}\ee
with $M_{br} $ treated here and in \cite{1} as a fitting
parameter, found from the pionless decay, and $\hat U(x)$ is a
chiral matrix $$\hat U (x) =\exp \left(i\gamma_5
\frac{\varphi_a\lambda_a}{f_\pi}\right);$$\be
\varphi_a\lambda_a\equiv \sqrt{2} \left(\begin{array}{lll}
\frac{\eta}{\sqrt{6}}+\frac{\pi^0}{\sqrt{2}},& \pi^+,& K^+\\
\pi^-,&\frac{\eta}{\sqrt{6}}-\frac{\pi^0}{\sqrt{2}},& K^0\\ K^-,&
\bar K_0, &-\frac{2\eta}{\sqrt{6}}\end{array}\right)\label{2}\ee

Note, that $\hat U(x)$ describes creation of any number of $\pi
(K, \eta)$ through small dimensional factor $f_\pi\cong 93$ MeV in
the denominator. This ensures strong  interaction and possible
multipion-quarkonuim resonances.

The $Q\bar Q$ Green's function with light quark loop inside and
with two possible ways of emission is shown in Fig. 1(a) and (b).
We shall keep here notations "a" and "b" for  the amplitudes with
subsequent one-pion and zero-pion--two-pion emissions,
respectively.

Using notations $n$ and $n'$ for $X(nS)$ and $X(n'S)$ states, and
$n_2, n_3$ the $ Q\bar q$ and $\bar Q q$ states, respectively, one
can write for  the amplitudes $a$ and $b$,
 \be
a_{nn'}=\gamma \sum_{n_2n_3} \int \frac{d^3p}{(2\pi)^3} \frac{
J^{(1)}_{nn_2n_3} (\vep, \vek_1) J^{*(1)}_{n'n_2n_3} (\vep,
\vek_2)}{E-E_{n_2n_3} (\vep) - E_\pi(\vek_1)} + (1\leftrightarrow
2)\label{3}\ee

$$
b_{nn'}=\gamma \sum_{n'_2n'_3} \int \frac{d^3p}{(2\pi)^3}
\left\{\frac{ J^{(2)}_{nn'_2n'_3} (\vep, \vek_1,\vek_2)
J^{*}_{n'n'_2n'_3} (\vep)}{E-E_{n'_2n'_3} (\vep) - E_\pi(\vek_1,
\vek_2)} +\right.$$
 \be \left.+\frac{ J_{nn^{\prime\prime }_2n^{\prime\prime }_3}(\vep)
J^{*(2)}_{n'n^{\prime\prime }_2n^{\prime\prime }_3} (\vep,
\vek_1,\vek_2)}{E-E_{n^{\prime\prime }_2n^{\prime\prime }_3}
(\vep)}\right\}.\label{4}\ee

Here $\gamma=\frac{M_{br}^2}{N_c}$, $J, J^{(1)}, J^{(2)}$ are
overlap integrals of heavy quarkonia wave functions $\psi_{Q\bar
Q}$ and the product of $Q\bar q$ and $\bar Q q$ wave functions
$\psi_{Q\bar q}\cdot $ $\psi_{\bar Q q}$ , multiplied by the
exponentials of free relative motion $\exp (i\vep\ver)$ with
momentum $\vep$, and  the pion plane waves with momenta $\vek_1$
and $\vek_2$.

\begin{figure}[!b]
\includegraphics[width=7cm,
keepaspectratio=true]{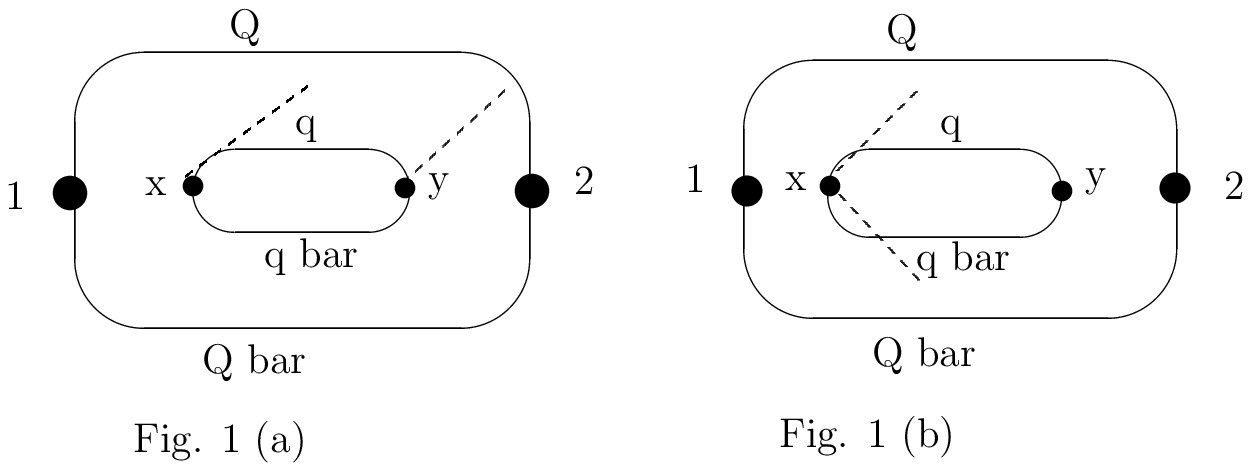}
\end{figure}

The resulting decay probability is \be \frac{dw}{dq} (n\to
n')\cong\frac{2}{\pi^3 N^2_c} \left( \frac{M_{br}}{f_\pi}\right)^4
\mu^2 \sqrt{x(1-x)} d \cos \theta |a-b|^2\label{5}\ee where $\mu^2
=(\Delta M)^2 -4m^2_\pi, ~~x= \frac{q^2-4m^2_\pi}{\mu^2},~~
q^2\equiv M^2_{\pi\pi} = (k_1+k_2)^2$ and $\theta$ is the angle of
emitted $\pi^+$ with  initial $X(nS)$ direction.
It is important, that $a$ and $b$ depend differently on $q,
\theta$: in $b$, the vectors  $\vek_1, \vek_2$ enter as a sum
$\veK=\vek_1+\vek_2$, while in $a$ one has a product of decreasing
functions of $|\vek_1|$ and $|\vek_2|$. E.g. in the Simple
Harmonic Oscillator (SHO) basis one can write  a form, exactly
satisfying the Adler zero condition: \be\begin{gathered}
\mathcal{M}\equiv a-b =const \left(
e^{-\frac{\vek^2_1+\vek^2_2}{4\beta^2_2}}\left(
\frac{1}{\tau_{nn'} (\omega_1)
+\omega_1}+\right.\right.\\+\left.\left.\frac{1}{\tau_{nn'}
(\omega_2) +\omega_2}\right)-
e^{-\frac{(\vek_1+\vek_2)^2}{4\beta^2_2}}\left(
\frac{1}{\tau_{nn'} (0)
}+\right.\right.\\+\left.\left.\frac{1}{\tau_{nn'}
(\omega_1+\omega_2)
+\omega_1+\omega_2}\right)\right)\end{gathered}\label{6}\ee


Here $\frac{1}{\tau+\omega} =\frac{1}{\lan \frac{\vep^2}{2\tilde
M}\ran +\omega}$ and $\beta_2$ is the SHO parameter for the $B,
B^*$ wave functions. Fitted to  known r.m.s. radius one has
$\beta_2 =0.5$ GeV. Now one can see, that
$\vek^2_1+\vek^2_2=\alpha(q) +\gamma(q) \cos ^2\theta$ is a weak
but increasing function of $q$, while $\veK^2$ is strongly
decreasing function of $q$. Expanding $a$, $b$ in powers of $x $
at the threshold, one arrives at the expression: \be \frac{dw}{dq}
=\frac{2}{\pi^3N^2_c} \left( \frac{M_{br}}{f_\pi}\right)^4
\frac{\mu^6}{(4\beta^2_2)^2} d\cos \theta \sqrt{x(1-x)}
b_{th}^2|\eta-x|^2\label{7}\ee

Here $\eta$ is a real parameter below $B\bar B$ threshold. We have
calculated the values of $\eta$ for $\Upsilon (nS)\to \Upsilon
(n'S)\pi\pi$, using the SHO basis with parameters fitted to all
r.m.s. radii of states and finding $\eta$ from (\ref{6}), where
$\tau_{nn'}$ were found from $\lan \frac{\vep^2}{2\tilde M}\ran$.
The intermediate states in all three transitions $(3\to 1),$ $
(2\to 1)$, and $(3\to 2)$ were chosen as $(B\bar B^*+ B^*\bar B)$
for the  amplitude $a$, and $(B\bar B)$ for the  amplitude $b$. We
have checked that the Adler zero condition is fulfilled when the
full set of intermediate states is involved, and imposed the Adler
zero condition in  our case of restricted set of intermediate
states, i.e. we have used the form (\ref{6}) with $a(k_1 =\omega_1
=0, k_2) = b (k_1=\omega_1 =0, k_2)$.

As a result we have obtained the following values of $\eta$:
$\eta(3\to 1) \cong0.39, ~~ \eta(2\to 1) =0.05$ and $\eta(3\to
2)=-3.2$.  On the other hand we have independently fitted the
three di-pion spectra measured by CLEO \cite{3} using the form \be
\frac{dw(n\to n')}{dq} =const. \sqrt{x(1-x)}|\eta_{fit}
-x|^2\label{8}\ee with constant and $\eta_{fit}$  as free
parameters and found that $\eta_{fit} (3\to 1) \approx 0.5, ~~\\
\eta_{fit} (2\to 1) =0, ~~\eta_{fit} (3\to 2) \approx -3$. Results
of this fitting are shown in Fig.2

3. The fitting curves in Fig.2 appear  to reproduce the main
features of di-pion spectra in all three cases. The deep well in
the $(3\to 1)$ double-peaked fitting curve is partially filled
since $\eta$ actually depends on $cos \theta$ and in integration
$\int d \cos \theta |\eta (\theta) -x|^2$ one has nonzero result
for all $x$.

It is interesting that the total yield of $\pi\pi$, calculated for
$\psi(2S) \to J/\psi \pi\pi$ with $M_{br} =1 $ GeV (fitted to
$\psi (3770)\to D\bar D)$, yields the value of $\Gamma_{\pi\pi}$
which is 3 times smaller than experiment. This is reasonable,
since $\Gamma_{\pi\pi}$ is proportional to the fourth power of the
overlap integral and the SHO wave function is certainly a very
rough approximation, nevertheless, the resulting $\Gamma_{\pi\pi}
$ is in the correct ballpark.

The di-pion spectrum for $\psi(2S) \to J/\psi\pi\pi$ also
corresponds to $\eta \approx 0$, while those of $\Upsilon (4S) \to
\Upsilon(n'S)_{\pi\pi} $ with $n'=1,2$ have $\eta_{fit}{(4S\to2S)}
=0.30$ and $\eta_{fit}(4S\to 2S)=0.61$ respectively. The
calculation of $\eta(4S\to n'S)$, however, should be done not with
SHO, but with realistic wave functions, which is in progress.





\begin{figure}[!b]
\vskip 0.5truecm \vspace{12pt}
 \hspace*{-0.2cm}
\includegraphics[width=10cm,height=12cm,keepaspectratio=true]{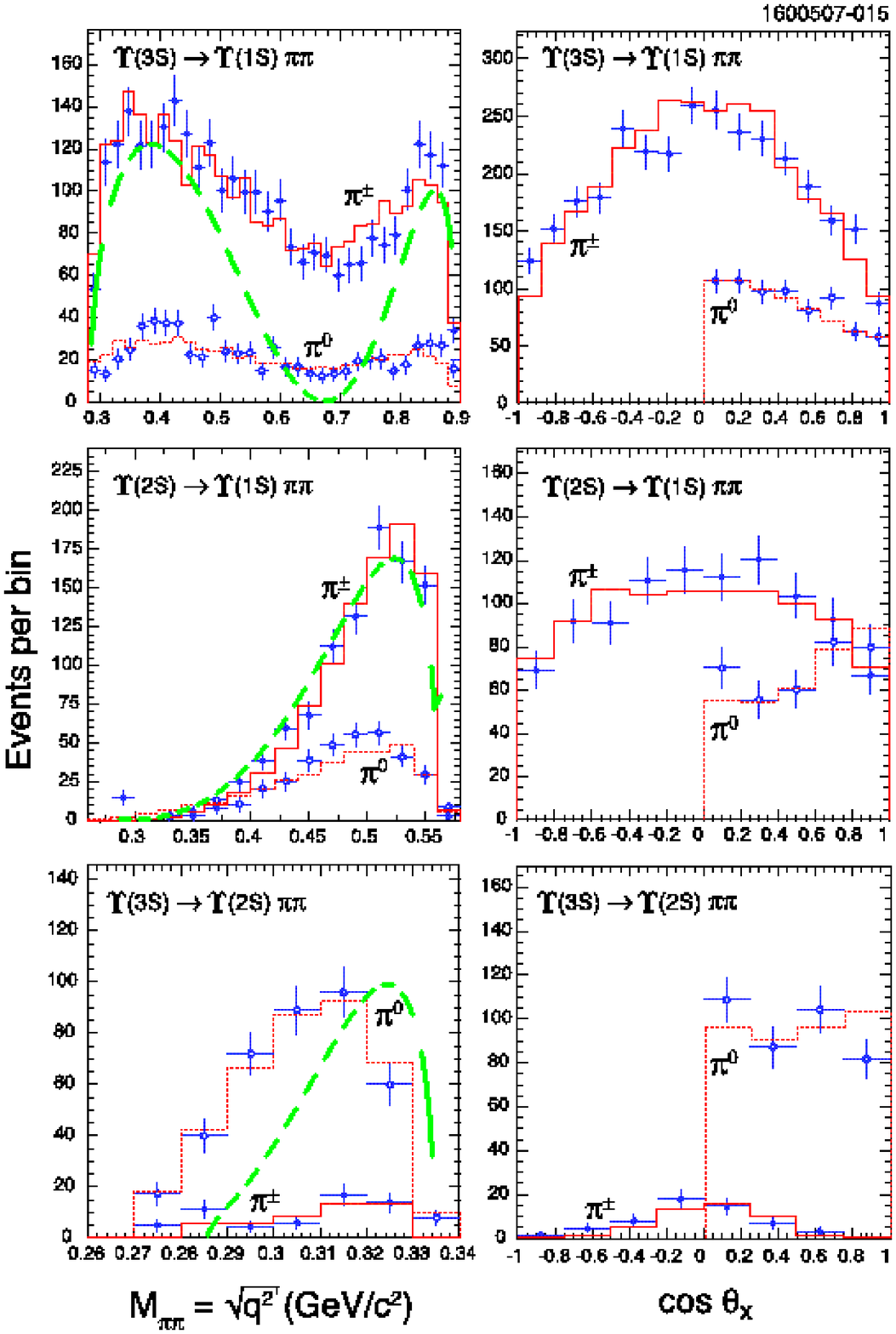}
\caption{}

{  Fig.2 The experimental data for $\Upsilon(nS)\to
\Upsilon(n'S)\pi\pi$ from \cite{3} versus theoretical
spectrum,Eq(\ref{7}) with $\eta=0.55; 0; -2.7$ (top to bottom,
dashed line).} \vspace*{2cm}
\end{figure}


 The processes  $\Upsilon (5S) \to \Upsilon
(n'S)\pi\pi$ with $n'=1,2$ were observed recently \cite{6} with
di-pion   spectra of possibly three-peaked form: also the total
di-pion yield $\Gamma_{\pi\pi}$ is thousand times larger, than for
$n=4S, 3S$. Both facts can be derived from our expressions, Eqs.
(\ref{4},\ref{6}), if one takes into account that the mass of
$\Upsilon(5S) =10.86$ GeV is above all three thresholds $B\bar B,
B\bar B^*, B^*\bar B^*$ and  the appearing imaginary part is very
large. This analysis is now in progress.

Finally, one can compare the total yield of $K^+ K^-$ to that of
$\pi^+\pi^-$ in $\Upsilon(5S\to 1S) $ transition. Indeed, the only
difference is in $\mu^2_i =(\Delta M)^2 -4 m^2_i,~~ i=\pi, K$.
From (\ref{7}) one can derive that $\Gamma_i= \int \frac{dw}{dq} d
q \sim \mu^7_i$, and $\frac{\Gamma_{K^+K^-} (5S\to
1S)}{\Gamma_{\pi\pi} (5S\to 1S)} \sim \left(
\frac{\mu_K}{\mu_\pi}\right)^7\approx 0.104$, which roughly agrees
with experimental data (an additional 50\% reduction of this ratio
follows if one use $f_K=0.112$ GeV$\neq f_\pi=0.093$ GeV).

Summarizing, we have suggested a new nonperturbative approach
which describes all transitions $X(n) \to X(n') \pi\pi (KK)$ for $
n=2,3$ and $n'=1,2$ without fitting parameters. It is  argued that
the standard MGFE approach to di-pion transitions  in heavy
quarkonia cannot be applied to large size  quarkonia.

 The author
is grateful to  S.I.Eidelman for stimulating  discussions and
useful suggestions and  for useful discussions to A.M.Badalian,
M.V.Danilov, A.B.Kaidalov, and all members of the ``Heavy
Quarkonia Workshop'' held  at ITEP, 28-29 November 2007.

   The financial support of RFFI grant 06-02-17012 and the grant for
   scientific schools NSh-843.2006.2 is gratefully acknowledged.

\end{document}